\documentclass[12pt]{article} 

\begin{document} 
\topmargin 0pt 
\oddsidemargin 0mm
\renewcommand{\thefootnote}{\fnsymbol{footnote}}
\begin{titlepage}
\vspace{5mm}

\begin{center}         
{\Large \bf  Newtonian Gravitomagnetism and analysis of Earth
  Satellite Results } \\
\end{center}
\vspace{6mm}
\begin{center}
{\large Harihar Behera\footnote{email: harihar@iopb.res.in}} \\
\vspace{5mm}
{\em
Dept. of Physics, Utkal University, Vanivihar, Bhubaneswar-751004,
Orissa,India}\\
 
\end{center}
\vspace{5mm}
\centerline{\bf {Abstract}}
\vspace{5mm}
                                              
  The possibility of a Newtonian gravitomagnetic field is considered here
with its immediate and far-reaching implications for the
interpretation of 2004 LAGEOS
experimental results confirming the general relativistic prediction of
Lense-Thirring effect.\\


\end{titlepage}
\section{Introduction}
The gravitomagnetic field \cite{1} is one of the most important
predictions
of general relativity, which is believed to have no Newtonian
counterpart \cite{1,2,3} , that emerges as a consequence of mass 
currents, analogous to the generation of magnetic field by electric
(charge) currents. Working within the framework of general
relativity, Lense and Thirring in 1918 \cite{4} predicted that a
spinning
massive body like Earth would generate a gravitomagnetic field,which
in turn would cause a precession of the orbits of its satellites. This 
effect, called Lense-Thirring (LT) effect ,is also known as ``dragging 
of inertial frames'' or more simply ``frame-dragging'' as Einstein
named it. Recently, frame-dragging or LT-effect is reported to have
been detected \cite {2,3,5} by Ciufolini et al. \cite {2,5} with
experimental results differing from the general relativistic
predictions by about $20\%$  during 1998-2002 \cite {2} and by$
10\%$ in 
2004 \cite {5}. Gravity probe-B ( an ongoing space mission of NASA to
test general relativity using cryogenic gyroscopes in orbit ) was 
launched in April 2004 and aims at measurement of Lense-Thirring effect
to about 1\% \cite {6} error. The important fact is that the analysis of the
 LAGEOS experimental results \cite {2,5} and the ongoing data analysis 
 of Gravity Probe B experiment \cite {6} are based on the assumption that 
 Newton's theory of gravitation has no phenomena analogous to
 magnetism although Newton's law of gravitation has a formal
 counterpart in Coulomb's law of electrostatics. In view of the above
 facts, the possibility of prediction of a Newtonian gravitomagnetic
 field is of immediate practical importance in the context of the
 gravitomagnetic effects are now becoming amenable to experimental
 observation. By noting the possible existence of a
 gravitomagnetic field first speculated by J. C. Maxwell in 1865 \cite 
 {7} and later pursued by Oliver Heaviside in 1893 \cite {8}, well
 before the advent of the relativity theories of Einstein, here we
 report the possibility of making a `natural' prediction of
 gravitomagnetic field within the framework of Newtonian Physics. The 
 full set of Faraday-Maxwell-type field equations of gravity
 describing Newtonian gravitodynamics has been derived here within the
 framework of Newtonian physics in a most natural way. The results
 represent a substantial advance in understanding Newton's theory of
 gravity in its entirety and also the important problem of the
 gravitomagnetic phenomena in physics. The conclusions seem to have
 immediate and far-reaching implications for the current research on
 the theory and experiments on gravitomagnetic effects in particular
 and gravitation in general.
\\
\section{History of the Gravitomagnetic field concept}
 Recognizing the striking formal analogy between Newton's law of
 gravitostatics and Coulomb's law of electrostatics, J. C. Maxwell
 \cite{7}, in one of his fundamental works on electromagnetism, turned
 his attention to the possibility of formulating the theory of gravity 
 in a form corresponding to the electromagnetic equations. In this
 attempt, Maxwell considered that the potential energy of a static
 gravitational configuration is always negative but he guessed that
 this should be re-expressible as an integral over field energy
 density, which being the square of the gravitational field ( by
 electromagnetic analogy ) is positive. Because of this puzzle,
 Maxwell did not work further on the topic. However, we will see a
 solution to Maxwell's puzzle in this work.    \\

Nonetheless, in 1893 O. Heaviside \cite {8} pursued Maxwell's attempt
further and wrote down the full set of Lorentz-Maxwell type equations
for gravity, very much analogous to the corresponding equations in classical
electrodynamics, by virtue of his power of speculative
thought. Heaviside's field equations implied the existence of
gravitational waves in vacuum, so he considered that the propagation
velocity of gravitational waves in vacuum might well be the speed of
light in vacuum. He also explained the propagation of energy in a
gravitational field, in terms of gravitoelectromagnetic Poynting
vector, even though he (as Maxwell did ) considered the nature of
gravitational energy is a mystery. Lacking experimental evidence of
gravitomagnetic effect and for some other reasons , he did not work
further. As per a report by McDonald (see McDonald in \cite{8}),
surprisingly Heaviside seemed to be unaware of the long history of
measurement of the precession of Mercury's orbit.  \\

The formal analogy was then explored by Einstein \cite{9}, in the
framework of General Relativity. Any theory that combines Newtonian
gravity together with Lorentz invariance in a consistent way, must
include a gravitomagnetic field, which is generated by mass
current. This is the case, of course, of General Relativity. May be
 Einstein had not seen Heaviside's
field equations when he was working on his relativistic theory of
gravity. Had Einstein seen Heaviside's field equations, his remark on
Newton's theory of gravity would have been different than what he made 
before the 1913 congress of natural scientists in Vienna \cite
{10},viz.,
\begin{quote}``After the untenability of the theory of action at distance had thus
been proved in the domain of electrodynamics,confidence in the
correctness of Newton's action-at-a-distance theory of gravitation was 
shaken. One had to believe that Newton's law of gravity could not
embrace the phenomena of gravity in their entirety, any more than
Coulomb's law of electrostatics embraced the theory of electromagnetic
processes.''\end{quote}

\section{Derivation of the field equations of Newtonian Gravitodynamics}
In Galileo-Newtonian physics, the source of gravitational field is
mass and mass is a conserved quantity - any flow of mass must come
from some supply, which we know to be true. The generation of
gravitational field is expressed by Gauss's law of gravitostatics :
\begin{equation}{\vec\nabla}\cdot\vec E_{g}\,=\, -4\pi G \rho 
\end{equation} 
where $\,\vec E_{g}\,$ is the gravitational field intensity,
$\,\rho\,$ (= mass density) is the source of $\,\vec E_{g}\,$ and $G$
is Newton's gravitational constant. The law of conservation of mass is 
expressed by the continuity equation :
\begin{equation}
\frac{ \partial \rho }{ \partial t}\,=\,-\,{\vec\nabla}\cdot\vec j
 \end{equation} 
where $\,\vec j\,$ is mass current density. Assuming the validity of
the two laws (1) and (2) ,we now proceed to find the gravitational
effects associated with mass currents. We know that there exist
physical systems where the two laws (1) and (2) are at play
simultaneously or co-work peacefully. To explore the gravitational
behavior of such systems, we take the time derivative of (1) to
obtain the equation
\begin{equation}{\vec\nabla}\cdot\left (\frac{1}{4\pi G}\frac{\partial\vec E_{g}}{\partial t}\,+\,\frac { \partial \rho }{ \partial
    t}\right )\,=\,0
\end{equation} 
Now by using (2) in (3),we obtain the equation :
\begin{equation}{\vec\nabla}\cdot\left (\frac{1}{4\pi G}\frac{\partial\vec E_{g}}{\partial t}\,-\,\vec j\right )\,=\,0
\end{equation} 
The quantity inside the parenthesis of (4) is a vector whose
divergence is zero. Since ${\vec\nabla}\cdot({\vec\nabla}\times{\vec
  A})\,=0 $ is true for any vector $\,\vec A\,$, the vector inside the 
parenthesis of (4) can be expressed as the curl of some other vector,
say $\,\vec H_{g}\,$. Mathematically speaking, the eq.(4) admits of
two solutions,viz.,
\begin{equation}{\vec\nabla}\times{\vec H_{g}}\,=\,\pm\,\left(\,-\,\vec
    j + \frac{1}{4\pi\,G}\frac { \partial \vec E_{g}}{ \partial
      t}\right ) 
\end{equation} 
which is formally analogous to Ampere-Maxwell's Law of classical
electrodynamics,viz.,
\begin{equation}{\vec\nabla}\times{\vec H}\,=\,\vec j_{e} +\,\frac {
    \partial \vec D}{ \partial t} \>\>\>\>{\rm ( in \ S. I. \ units )}
\end{equation}
where $\,\vec H\,$ is the magnetic field, $\,\vec j_{e}\,$ is electric 
current density and $\,\vec D\,$ is the displacement vector. In empty
space $\,\vec D\,=\,{\epsilon_{0}}\vec E\,$, where $\,\epsilon_{0}$ is 
the permitivity of empty space and $\,\vec E\,$ is the electric field.
One of the solutions (5) would correspond to the reality and the other 
might be a mathematical possibility having no or new physical
significance in which we are not presently interested in. For our
present purpose, we have to choose one of these two solutions that may 
correspond to the reality. Which one to choose ? We can answer this
question by following the rule of `study by analogy'. To follow this
rule, let us write down the Gauss's law of electrostatics in the form
:
\begin{equation}{\vec\nabla}\cdot\vec E\,=\,\frac {\rho_{e}}{\epsilon_{0}}
\end{equation} 
where $\,\rho_{e}\,$ is the charge density, $\,\vec E\,$ is electric
field intensity,$\epsilon_{0}$ is the permitivity of empty
space. Eq.(7) is analogous to Eq.(1),with only a difference in the sign
of the source functions. From (6) and (7) we find that $\rho_{e}$ and
$\vec j_{e}$ have got the same sign. So by analogy we infer that in
case of gravity $\rho$ and $\,\vec j\,$ should have the same sign and
the signs of $\,\vec j_{e}\,$ and $\,\vec j\,$ should be opposite. So
by this logic, our most natural choice for the real solution of (4) is
\begin{equation}{\vec\nabla}\times{\vec H_{g}}\,=\,-\,\vec j +
  \frac{1}{4\pi\,G}\frac { \partial \vec E_{g}}{ \partial t} 
\end{equation} 
The comparisons of (1) with (7) and (6) with (8) suggest us to
introduce or define the following terms in gravitation,viz.,$\,\vec
E_{g}\,$ as the gravitoelectric (GE) field, $\,\epsilon_{0g}\,=\,\frac 
{1}{4\pi\,G}\,$ as the gravitoelectric permitivity of empty space,
$\,\vec D_{g}\,=\,\epsilon_{0g}\vec E_{g}\,$ as the gravitational
displacement vector in empty space, $\,\vec H_{g}\,$ as the
gravitomagnetic field, and name the Eq.(8) as the gravitational
Ampere-Maxwell Law of Newtonian gravitoelectromagnetic (GEM) theory or 
Newtonian gravitodynamics. The appearance of the sign difference of
the corresponding source functions in (1)and (7),(6) and (8) implies the
characteristic dissimilarity between the two fundamental interactions
of nature - which are opposite in nature. For instance, in
electromagnetism like charges repel and unlike charges attract under
static conditions, but under dynamic condition the nature of the
interaction gets reversed - like electric currents (i.e. parallel
currents) attract and unlike (i.e. anti-parallel) currents repel. In
case of gravitation, we have the opposite situation, viz.,like masses
attract [ and if negative mass \cite{11} exists,the unlike masses
should repel ] under static condition, and by the nature analogy
between gravitational and electrical phenomena, we would have a
reversed situation in the dynamic condition,viz., like (i.e. parallel) 
mass currents would repel ( as a form of anti-gravity \cite{12}) and
unlike (i.e. anti-parallel) mass currents would attract. The deep
analogy between electrical and gravitational phenomena in the frame
work of general relativity has earlier been discussed by
R. L. Forword\cite{12} , R. Wald \cite{13},
V. B. Braginsky,C.M. Caves and K.S. Thorne \cite{14}. One of the
implications or predictions of gravitational Ampere-Maxwell Law (8) is 
that mass currents would generate gravitomagnetic field in accordance
with what we may call the gravitational Ampere's Law of gravitomagnetostatics,viz.,

\begin{equation}{\vec\nabla}\times{\vec H_{g}}\,=\,-\,\vec
    j \,\,\,\,\,\,\,\,\,\,\,\,\,\,\,\,(\,{\rm when \,\,\,\frac{\partial \vec E_{g}}{ \partial
      t}\,=\,0\,\,\,\, in\,\,\, (8)\,} )
\end{equation} 
which is the gravitational analogue of Ampere's Law in magnetostatics,viz.,
\begin{equation}{\vec\nabla}\times{\vec H}\,=\,\vec j_{e} \,\,\,\,\,\,\,\,\,\,\,\,\,\,\,\,(\,{\rm when\,\,\,\frac{\partial \vec D}{ \partial t}\,=\,0\,\,\,\, in\,\,\, (6)\,} )
\end{equation} 
To understand the significance and the implications of the 2nd term on 
the right hand side of (8),let us note the significance and the
implications of the term $\,\frac{\partial \vec D}{ \partial
  t}\,(=\,\epsilon_{0}\frac{\partial \vec E}{ \partial t})\,$ in
Ampere-Maxwell's Law (6). This term was added to Ampere's Law (10) by
Maxwell who called it the displacement current. The displacement
current is of crucial importance for rapidly fluctuating
fields. Without it there would be no electromagnetic wave
\cite{15}. Analogously the gravitational displacement current $\,\frac{\partial \vec D_{g}}{ \partial
  t}\,(=\,\epsilon_{0g}\frac{\partial \vec E_{g}}{ \partial t})\,$ is
of crucial importance for rapidly fluctuating gravitational
fields. Without it there would be no gravitational wave. Because of
these implications of gravitational Ampere-Maxwell Law (8), it is
natural to assume the existence of gravitational waves in empty space
where the law still stands even if $\,\vec j\,=\,0\,$. Thus if
gravitational waves exist in empty space, then the electromagnetic
analogy we are uncovering suggests that the wave equation for the
fields $\,\vec E_{g}\,$ and $\vec H_{g}\,$ in empty space should stand 
as
\begin{equation}
\nabla^{2}\cdot\vec H_{g}\,-\,\frac{1}{{c_{g}}^{2}}\cdot\frac{\partial^{2}\vec H_{g}}{\partial t^{2}}\,=\,0  
\end{equation}
\begin{equation}
\nabla^{2}\cdot\vec E_{g}\,-\,\frac{1}{{c_{g}}^{2}}\cdot\frac{\partial^{2}\vec E_{g}}{\partial t^{2}}\,=\,0  
\end{equation}
where $\,c_{g}\,$ is some finite speed of propagation of gravitational 
waves, which is to be determined by experiments. To explore this
possibility, let us take the curl of (8) to obtain the equation
\begin{equation}
\nabla^{2}\vec H_{g}\,+\,\frac{1}{4\pi\,G} \frac{\partial }{\partial
  t} (\vec{\nabla}\times{\vec E_{g}})\,=\,\vec{\nabla}\times{\vec
  j}\,+\,\vec{\nabla}\cdot(\,\vec{\nabla}\cdot{\vec H_{g}}) 
\end{equation}
In empty space (13) reduces to the wave equation (11), provided the
following relations
\begin{equation}
\vec{\nabla}\times{\vec E_{g}}\,=\,-\,\frac
{4\pi\,G}{{c_{g}}^{2}}\frac{\partial\vec H_{g}}{\partial
  t}
\end{equation}
\begin{equation}
\vec{\nabla}\cdot{\vec H_{g}}\,=\,0
\end{equation}
\begin{equation}
\vec{\nabla}\times{\vec j}\,=\,0
\end{equation}
hold good in empty space. The relation (16) is certainly true in empty 
space where $\,\vec j\,=\,0\,$. But the relations (14) and (15) have
to be accepted if the existence of gravitational waves in empty space
is accepted. We now recognize that we have arrived at the full set of
Faraday-Maxwell-type field equations of gravity which may be
represented by
 \begin{equation}
\vec{\nabla}\cdot{\vec{E}_{g}} = -4\pi G \rho =
-{\rho}/{\epsilon_{0g}},\; \;\;\;\;\;\; \; \; {\rm {by\,\,\, defining \;\;\;\;  \epsilon_{0g}={1}/4\pi G}}
\end{equation}
\begin{equation}
\vec{\nabla} \times \vec{B}_g = - \mu_{0g} \vec j + (1/ {c_{g}}^{2})
({\partial \vec{E}_g }/{\partial t}) ,\;\;\;\;\,\,\, {\rm{\,\,by\,\,defining\;\;\;\; \mu_{0 g}}}
= {4{\pi}G}/{c_{g}}^{2} 
\end{equation}
\begin{equation}
\vec{\nabla} \cdot \vec{B}_{g} = 0
\end{equation}
\begin{equation}
\vec{\nabla} \times \vec{E}_g = - \partial \vec{B}_g / \partial t
\end{equation}
where we have defined $\,\vec B_{g}\,=\,({4{\pi}G}/{c_{g}}^{2})\vec
H_{g}\,=\,\mu_{0 g}\vec H_{g}\,$ as the gravitomagnetic induction
field in empty space and $\mu_{0 g}$ as the gravitomagnetic
permeability of empty space, in analogy with the electromagnetic case
where magnetic induction $\,\vec B\,$ in empty space is defined by
$\,\vec B\,=\,\mu_{0}\vec H\,$, $\,\mu_{0}\,$ is the magnetic
permeability of empty space. This definition of $\,\mu_{0g}\,$ ensures 
the relation $\,c_{g}\,=\,{(\epsilon_{0g}\mu_{0g})}^{-1/2}$ in complete 
analogy with its electromagnetic counterpart :$\,c\,=\,
{(\epsilon_{0}\mu_{0})}^{-1/2}$ . As in electrodynamics, $\,\mu_{0g}\,$ 
would be the coefficient that would determine the strength of
gravitomagnetic interaction or effects. Assuming $\,c_{g}\,=\,c\,$, a
typical value of $\,\mu_{0g}\,$ may be estimated at
$\,\mu_{0g}\,=\,9.33\times{10}^{-27} N.{s}^{2}/{kg}^{2}$. This gives us 
a glimpse of the order of magnitude of $\,\mu_{0g}\,$, for we know not
yet the exact value of $\,c_{g}\,$ to calculate $\,\mu_{0g}\,$. It is this order of smallness
of the the value of $\,\mu_{0g}\,$ that makes the strength of
gravitomagnetic effects or interaction very negligible. From (18) it is 
now clear that very large mass currents or very rapidly fluctuating
fields are required for production of gravitomagnetic field $\,\vec
B_{g}\,$ of  appreciable strength. Therefore any search for
gravitomagnetic effects should search for physical systems or
processes where the mass current density is very large or the gravitational 
field fluctuation is very rapid. Such conditions or situations are
available in certain astrophysical systems or events. The smallness of 
the value of $\,\mu_{0g}\,$ also explains why the existence of 
gravitomagnetic effects had escaped the human detection in spite of
the long history of its study. By the way, it is to be noted that in
1893 Heaviside \cite {8} speculated exactly these equations (17-20) with
different notations and definitions of the various quantities
involved.\\
To complete the dynamical picture we ask : What replaces the equation
$\,\vec F\,=\,m\vec E_{g} \,$ to describe the force on a particle of
mass $\,m\,$, when that particle moves with some velocity $\,\vec v\,$ 
in given gravitoelectric and gravitomagnetic fields $\,\vec E_{g}\,$
and $\,\vec B_{g}\,$ ? Because of the rich and detailed
correspondence between electrical and gravitomagnetic phenomena
uncovered above, one may at once suggest ( as Heaviside did ) the
force on mass $\,m\,$ is now 
\begin{equation}
\vec F\,=\,m\vec E_{g}\,+\,m\vec v\times\vec B_{g}
\end{equation}
where $\,\vec v\,$ is the velocity of mass $\,m\,$ in this expression, 
in analogy with the Lorentz force Law of electrodynamics :
\begin{equation}
\vec F\,=\,q\vec E\,+\,q\vec v\times\vec B
\end{equation}
where the symbols have their usual meanings. It is to be noted that
the gravitational Lorentz force law (21) can also be derived from
other Newtonian assumptions chosen by Feynman and Dyson \cite {16} and 
following the path of Feynman's proof of Mawwell's equations as
reported by Dyson \cite {16} with his editorial comment on the
derivations. The other way to arrive at the above Lorentz-Maxwell-type 
equations  of gravity may be that of the Newtonian way Schwinger
 et al.\cite{17} have chosen to infer the Lorentz-Maxwell equations of classical
electrodynamics. From all these variant approaches which lead to the
same equations as detailed above, we are convinced that the set of four
equations (17)-(20) would form the basis of all Newtonian
gravitoelectromagnetic phenomena . When combined with the
gravitational
Lorentz force equation (21) and Newton's 2nd law of motion, these
equations would provide a complete description of the dynamics of the
interacting massive particles and gravitomagnetic fields.\\

To compare the field equations (17)-(20) with those predicted by
general relativity (GR), we note the following approximations to the
Maxwell-type field equations of GR in the parametrized-post-Newtonian
(PPN) formalism \cite {14},which as per the present definitions and
notations may be written as 
\begin{equation}{\vec\nabla}\cdot\vec E_{g}\,\cong\, -4\pi G \rho 
\end{equation} 
\begin{equation}\vec\nabla\times{\vec E_{g}}\,=\,-\,
\frac{ \partial \vec B_{g}}{ \partial t} 
\end{equation} 
\begin{equation}{\vec\nabla}\cdot{\vec B_{g}}\,=\,0 
\end{equation} 
\begin{equation}{\vec\nabla}\times{\vec B_{g}}\,=\,\left({\frac{7}{2}}\Delta_{1} + {\frac{1}{2}}\Delta_{2} \right)\left( - \frac{4\pi  G}{c^{2}}\rho\vec v + \frac{1}{c^{2}}\cdot\frac { \partial \vec E_{g}}{ \partial t}\right) 
\end{equation}
where $ \Delta_{1} $ and $ \Delta_{2} $ are PPN parameters, $ \rho
$ is the density of rest masses in the local frame of the matter, $
\vec v $ is the ordinary  (co-ordinate) velocity of the rest mass
 relative to the PPN co-ordinate frame. In general relativity  $ \left(\frac{7}{2}\Delta_{1} + \frac{1}{2}\Delta_{2}\right)\,\cong\, 4 $  and so Eq.(26) can be rewritten as
\begin{equation}{\vec\nabla}\times{\vec B_{g}}\,\cong\,-\,\frac{16\pi  G}{c^{2}}\rho\vec v + \frac{4}{c^{2}}\cdot\frac { \partial \vec E_{g}}{ \partial t} 
\end{equation} 
In empty space (where $ \rho = 0 $), these field equations reduce 
to the following equations: 
\begin{equation}{\vec\nabla}\cdot\vec E_{g} = 0 
\end{equation}
\begin{equation}{\vec\nabla}\times{\vec E_{g}} = -\,\frac{ \partial\vec B_{g}}{ \partial t} 
\end{equation}
\begin{equation}{\vec\nabla}\cdot{\vec B_{g}} = 0
\end{equation}                                                                 \begin{equation}{\vec\nabla}\times{\vec B_{g}} =  \frac{4}{c^{2}}\cdot\frac { \partial \vec E_{g}}{ \partial t} 
\end{equation}
As per our previous analysis, these Maxwell-type equations of GR imply
that
\begin{equation}\epsilon_{0g,GR}\,=\,1/4\pi\,G,\,\,\,\,\,\,\,\mu_{0g,GR}\,=\,16\pi\,G/c^{2}, 
  \end{equation}
  which in turn imply that in the low velocity and weak field
  approximation of GR the speed of gravitational wave in empty space
  is $\,c_{g,\,GR}\,=\,{(\epsilon_{0g,GR}\mu_{0g,GR})}^{-1/2}\,=\, c/2\,$, a result not expected from a Lorentz
  covariant theory of gravitation. This value of $\,c_{g}\,=\,c/2\,$
  can also be inferred from the wave equations that follow from the
  above equations (28)-(31) by taking the curl of (29) and utilizing Eqs.(28) and (31) we get the wave equation for the field  $ \vec E_{g} $ in empty space as 
\begin{equation}{\vec\nabla^{2}}\cdot\vec E_{g} -  \frac{1}{{c_{g}}^{2}}\cdot{\frac{\partial^{2}\vec E_{g}}{\partial t^{2}}}  = 0 
\end{equation}
where  $ c_{g} = c/2 $. Similarly the wave equation for the field  $
\vec B_{g} $ can be obtained  by taking the curl of Eq.(31) and
utilizing Eqs.(29) and (30) : 
\begin{equation}{\vec\nabla^ {2}}\cdot\vec B_{g} - \frac{1}{{c_{g}}^{2}}\cdot\frac{\partial^{2}\vec B_{g}}{\partial t^{2}} = 0
\end{equation} 
where  again we get  $ c_{g}  = c/2$ . However, Peng in \cite {18} has 
discussed a set of Maxwell-like equations that arise in the slow
motion, weak field limits of Einstein's field equations which in the
present notation agrees with Newtonian gravitodynamic equations
(17-20) with
$\,c_{g}\,=\,c\,$ \cite {19,20}. In \cite {20} Peng's equations have
been utilized to predict the quantization of planetary revolution
speeds which matches with the experimental data. It is to be noted that Ciufolini's theoretical analysis
of LAGEOS data is based on the gravitomagnetic field $\,\vec B_{g}\,$ that can be
obtained from Eq.(27) which differs from Peng's corresponding equation 
by a factor of 4 \cite {18,19,20} which is responsible for the above
result $ c_{g}  = c/2$ .\\

\section{Immediate theoretical consequences of Interest}
 Newtonian gravitodynamics is very much
analogous to Maxwell's electrodynamics as revealed by the form of its equations. Therefore gravitational
phenomena very much analogous to those of electromagnetic theory are
not surprising to be revealed by this theory. However few concepts and
results of unconventional nature and importance may be discussed as
under. 
Historians tell us that Newton was quite unhappy over the fact that his 
law of gravitation implies an action-at-a-distance interaction over
very large distances such as that between the Sun and the Earth. But he 
was unable to resolve this problem \cite {21}. To resolve  Newton's 
problem within the framework of Newtonian physics and in terms of
potential functions of Newtonian gravitodynamics, we note that the
homogeneous equations (19) and (20) admit of the solutions :
\begin{equation}\vec B_{g} = \,\vec{\nabla}\times{\vec A_{g}},\>\>\>\>\>\>\>\>  \vec E_{g} = -\,\vec{\nabla }\cdot \Phi_{g} - {\partial{\vec A_{g}}/{\partial t}}
\end{equation}
where ${ \Phi_{g}} $ and ${ \vec A_{g}} $ represents respectively the
gravitational scalar and vector potential of this  theory.These
potentials satisfy the inhomogeneous wave equations :
\begin{equation}\nabla^{2}\cdot\Phi_{g}\, - \,\frac{1}{{c_{g}}^{2}}\cdot\frac{\partial^{2}\Phi_{g}}{\partial t^{2}}\,=\,4\pi\,G\, \rho\,=\,\rho/\epsilon_ {0\,g}  
\end{equation}
\begin{equation}\nabla^ {2}\cdot\vec A_{g}\, -\,
  \frac{1}{{c_{g}}^{2}}\cdot\frac{\partial^{2}\vec A_{g}}{\partial
    t^{2}}\,=\,\frac{4\pi\,G}{c_{g}^{2}}\vec j\,= \,\mu_{0\,g}\vec j 
\end{equation}
if the gravitational Lorenz \cite {22} gauge condition
\begin{equation}\vec{\nabla}\cdot\vec
  A_{g}\,+\,\frac{1}{{c_{g}}^{2}}\frac{\partial\Phi_{g}}{\partial{t}}\,=\,0
\end{equation}
is imposed. These will determine the generation of gravitational waves 
by prescribed mass and mass current
distributions. Particular solutions (in vacuum) are 
\begin{equation}\Phi_{g}\,(\,\vec r\,,t\,)\,=\,-\,G\,\int\,{\frac{\rho(\,\vec r^{\prime}\,,\,t^{\prime}\,)}{\,|\vec r\,-\,\vec r^{\prime}\,|}dv^{\prime}}
\end{equation} 
\begin{equation}\vec A_{g}\,(\,\vec
  r\,,t\,)\,=\,-\,\frac{G}{{c_{g}}^{2}}\,\int\,{\frac{\vec j(\,\vec r^{\prime}\,,\,t^{\prime}\,)}{\,|\vec r\,-\,\vec r^{\prime}\,|}dv^{\prime}}
\end{equation}
 where $\,t^{\prime}\,= t\,-\,{|\,\vec r\,-\,\vec
   r^{\prime}\,|}/c_{g}\,$  is the retarded time. These are called the
 retarded potentials. Thus we saw that retardation in Newtonian gravity is
 possible in flat space and time in the same procedure  as we adopt
 in electrodynamics. Had Einstein seen these possibilities, his
 confidence in the correctness of Newton's theory would not have been
 shaken and his approach to relativistic gravity would have been
 different with the implication that the history and the
character of gravitation might have been different from what we know
today.
Now let us come to Maxwell's puzzle over the gravitational field
energy density. Actually this puzzle is not a real puzzle, but a guess-work of Maxwell, since the actual calculation done by electromagnetic procedure
yields the energy density of gravito-electric
 (i.e.the electric-type component of gravity) and gravitomagnetic
 (i.e. the magnetic-type component ) field , respectively as 
\begin{equation}
  (i)\,\,\,u_ {g\,e}\,=\,-\,{\frac{1}{2}} \epsilon_ {0\,g} \vec E_ {g} \cdot 
  \vec E_ {g},\,\,\,\,\,\,\,(ii)\,\,\,u_ {g\,m}\, = \,-\,\frac {1}{2
    \mu_ {0\,g}}\, \vec B_ {g}\cdot \vec B_ {g} 
\end{equation} 
where  $\, \epsilon_{0\,g}\,=\,{1}/{4\,\pi\,G\,}$ and
  $\,\mu_{0\,g}\,=\,4\,\pi\,G/{c_{g}}^{2}\,$ and the total field energy
  density is given by a sum of the above two,i.e.
\begin{equation}u_ {f\,i\,e\,l\,d}\,\,=\,\,u_ {g\,e}\,+\,u_ {g\,m} 
\end{equation} 
For a particle at rest, i.e., in gravitostatics, the only contribution
to its gravitational field energy is that due to the gravito-electric
field. In
gravitostatics ,it is easy to compute the gravitational or gravito-electric
self energy of a sphere of radius $ R $ and mass $\,M\,$ with
uniform mass density by using Eq.(41i),which comes out as : 
\begin{equation} U_{g}\,=\,-\,{\frac{1}{2}}\,\epsilon_ {0\,g}{\int_{0}^\infty}{E_ {g}}^{2}\,4 \pi r^{2}\,dr\,\,=\,-\,\frac{3G{M}^{2}}{5R} 
\end{equation} 
The result (43) is in complete agreement with the Newtonian
result. It is to be noted that Visser \cite {23} used  exactly this
definition of gravitational field energy density in his classical
model for the electron. Such a definition of the field energy of
gravity has the  advantage of describing the correct nature of
gravitation on quantization because in analogy with electromagnetic
theory, the present theory will eventually lead to a gauge  field  of
spin $1$ and the spin $1$ gauge fields having positive and definite field
energy, on quantization, as we know lead to a repulsive force field
for identical charges of such fields. It is due to this reason Gupta
\cite {24} and Feynman \cite {25}
suggested  rejection of any spin $1$ gauge theory of gravity with field
energy being positive and definite as such fields do not account  for
the observed nature of gravitational interaction.Since general
relativity speaks of spin $2$ tensor theory of gravity and Newtonian
theory as developed here implying spin $1$ vector theory of gravity
very much analogous to the electromagnetic theory, one of the
fundamental questions raised in the Dicke Framework (see in \cite {26}) :
\begin{quote} `` What types of fields, if any, are associated with
  gravitation - scalar fields, vector fields, tensor fields, ...........?''\end{quote}requires immediate attention to understand the riddle of gravitation 
  better. The very possibility of a Newtonian gravitoelectromegnetic
  (GEM) theory also raises another important question : `` Does
  general relativity obey the correspondence principle ?'' When it
  does, (as per a comparison of Peng's \cite {18} work with the
  present one assuming $\,c_{g}\,=\,c\,$ )  do the predictions of gravitomagnetic phenomena agree with experimental results \cite{2,5}?\\
\section{Analysis of Earth Satellite (LAGEOS) Data}
As an immediate application of Newtonian gravitodynamics, let us consider the analysis
of Earth satellite (LAGEOS and LAGEOS2) data \cite {2,5}.In the slow motion and
weak field limit of GR, the
gravitomagnetic induction field $\,\vec B_{g,GR}\,$ associated with
the spin-angular momentum $\,\vec J\,$ of a spherical spinning body (
such as the Earth ) can be
estimated from (27) following the standard electromagnetic procedure
of estimation of magnetic field generated by localized current
distributions \cite{15}. The result is that
\begin{equation} \vec B_{g ,GR} = 2G[\vec Jr^{2} -3( { \vec J \cdot \vec{r} }) {\cdot \vec r} ]/(c^{2}r^{5})
\end{equation}
where $\,\vec r\,$ is the position vector of the field point from the
center of the Earth. For a satellite orbiting the Earth having a
gravitomagnetic field presumed at that given by (44), the
Lense-Thirring (LT) nodal precession is estimated \cite {1} at
\begin{equation}
{\dot{\vec{\Omega}}_{LT}}^{GR}\,=\,\frac{2G\vec{J}}{c^{2}a^{3}(1-e^{2})^{3/2}}
\end{equation}
where $\,a\,$ and $\,e\,$ respectively represents the semi-major axis
and the eccentricity of the satellite orbit. In 2004 Ciufolini and
Pavlis \cite {5} measured the Lense-Thirring nodal precession at 
\begin{equation}
\dot{\vec\Omega}_{LT}^{experiment}\,=\,(0\cdot99\,\pm0\cdot10)\dot{\vec\Omega}_{LT}^{GR}
\end{equation}
But the above interpretation has not taken care of the contribution arising
out of the Newtonian gravitomagnetic field, which as we saw here a
theoretical possibility. By the same procedure,one can estimate the
Newtonian gravitomagnetic field of Earth at
\begin{equation} \vec B_{g ,New} =\,G[\vec Jr^{2} -3( { \vec J \cdot \vec{r} }) {\cdot \vec r} ]/(2{c_{g}^{2}}r^{5})
\end{equation} 
and this field would then cause an LT-type nodal precession at
\begin{equation}
{\dot{\vec{\Omega}}_{LT}}^{New}\,=\,\frac{G\vec{J}}{2{c_{g}}^{2}a^{3}(1-e^{2})^{3/2}}\,=\,{\left(\frac{c}{2c_{g}}\right)}^{2}\,\dot{\vec\Omega}_{LT}^{GR}
\end{equation}
To see the implications of (48) we consider the following two cases of
 interest,viz., \\
(a) if $\,c_{g}\,=\,c\,$, then
$\,{\dot{\vec{\Omega}}_{LT}}^{New}\,=\,\frac{1}{4}\,{\dot{\vec\Omega}}_{LT}^{GR}\,=\,25\% 
$ of ${\dot{\vec\Omega}}_{LT}^{GR}\,$,and\\
(b) if $\,c_{g}\,=\,c/2\,$, then
$\,{\dot{\vec{\Omega}}_{LT}}^{New}\,=\,{\dot{\vec\Omega}}_{LT}^{GR}\,=\,100\% 
$ of ${\dot{\vec\Omega}}_{LT}^{GR}\,$.\\
It is to be noted that the exact value of $c_{g}$ in the slow motion
and weak field approximation of GR is not unique as seen in sec.3 of
this paper. Because of the uncertainties in the exact value of $c_{g}$ 
and the possible existence of a Newtonian gravitomagnetic field, the
interpretation of the LAGEOS experimental result \cite{5} as a pure
general relativistic effect with $99\%$ claimed accuracy of GR is
doubtful. It may be noted that the 2004 claim of Ciufolini and Pavlis
\cite{5} had been earlier suspected by Iorio \cite{27,28} from other possible
sources of gravitational error which of course have already been
addressed in the authors' 2005 work \cite{5} as completely
unfounded. Nonetheless,Iorio \cite{29} still suspects the correctness
of the latest results \cite{5}. However the most likely Newtonian gravitomagnetic error 
introduced by the result (a) considered above for the case $c_{g}=c$ 
seems to be large enough to be of some concern let alone the possibility
$c_{g}=c/2$ leading to the result (b) which appears to be  more
serious one .\\

\section{Concluding Remarks}
  The possibility of a Newtonian gravitomagnetic field can not be ruled 
  out in principle and therefore should be taken care of in the current 
  and future measurements of gravitomagnetic phenomena. This
  possibility adds to our understanding of Newtonian gravity in its
  entirety and raises some fundamental questions of importance in the
description of gravitational phenomena which may deserve certain
attention. The analysis and the interpretation of the earth satellite
(LAGEOS and LAGEOS2) data \cite {2,5} which ignore(s) the Newtonian 
gravitomagnetic contribution to the observed effects may be 
reconsidered or re-examined for an unambiguous explanation of the
experimental observation.

\textbf{Acknowledgments}
The author is very much grateful to Prof.Lorenzo Iorio, Viale Unita di
Italia, Bari, Italy for his interest in this work, for poviding some 
valuable references on the topic, for encouraging and valuable
suggestions and for permitting the author to mention him in the 
acknowledgements. The author is also very much indebted to
Prof. N. Barik, Prof. L. P. Singh, Prof. N. C. Mishra, Prof. S. Jena,
Dr. P. Khare, Dr. K. Moharana all of Department of Physics, Utkal
University, Bhubaneswer, India and N. K. Behera of U. N. College, Soro
 Balasore, India, for discussins and helpful suggestions. The help
 received from the Institute of Physics, Bhubaneswar for using the
 library and computer facility for this work is gratefully
 acknowledged by the author.\\

                
\end{document}